\documentclass[12pt]{iopart}
\usepackage{iopams}
\usepackage{amssymb,bm,verbatim}
\newcommand{\E}{\mathcal{E}}
\newcommand{\w}[1]{\bm{#1}} 
\def\ud{\textrm{d}}

\begin{document}

\title[Shearfree perfect fluids]{Shear-free perfect fluids with a solenoidal electric curvature
}
\author{Norbert Van den Bergh$^1$, John Carminati$^2$, Hamid Reza Karimian$^{1,3}$ and Peter Huf$^2$}
\address{$^1$\ Faculty of Applied Sciences FEA16, Gent University, Galglaan 2, 9000 Gent, Belgium}
\address{$^2$\ School of Engineering and Information Technology, Deakin University, Australia}
\address{$^3$\ Faculty of Sciences, Ilam University, Banganjab, Ilam, Iran}
\eads{\mailto{norbert.vandenbergh@ugent.be}, \mailto{jcarm@deakin.edu.au},\\
\mailto{Hamidreza.Karimian@UGent.be}}

\begin{abstract}
We prove that the vorticity or the expansion vanishes for any shear-free perfect fluid
solution of the Einstein field equations where the pressure
satisfies a barotropic equation of state and the spatial divergence
of the electric part of the Weyl tensor is zero.
\end{abstract}


\section{Introduction}
There has been much effort devoted to establishing the \emph{shear-free fluid conjecture}, that
general relativistic, shear-free perfect fluids which obey a barotropic
equation of state $p = p(\mu)$ such that $p + \mu \neq  0$, are either non-expanding
($\theta= 0$) or non-rotating ($\omega= 0$). 
This conjecture appears as such for the first time in Treciokas and Ellis\cite{rad} and
has been established in many
special cases, but a general proof or counter-example is still lacking. In
support, the conjecture is known to hold in cases where: $p = const$ (dust
with a cosmological constant)\cite{dust1,dust2,dust3}; spatial homogeneity\cite{spat1,spat2};
$\ud p/\ud \mu = 1/3$ (incoherent radiation)\cite{rad}; $\ud p/\ud \mu = -1/3$\cite{langth,collinp} or $1/9$\cite{39state2}; 
$\w{\omega}$ and $\dot{\w{u}}$ are parallel\cite{collinsKV}; vanishing of the magnetic part $\w{H}$\cite{collinsm} or of the electric part 
$\w{E}$\cite{langth,carme} of the Weyl tensor; $\theta=\theta(\mu)$\cite{langcollins} or $\theta=\theta(\omega)$\cite{carme}; 
Petrov types N\cite{carmn} and III\cite{carm3a,carm3b}; there
exists a conformal Killing vector parallel to the fluid flow $\w{u}$\cite{coley}. 
It is noteworthy that there exist Newtonian perfect fluids with a barotropic equation of state, which are rotating, expanding but non-shearing. Hence, if
true, such behaviour of fluids would be a purely relativistic effect. Recently, in an attempt to generalize the result established by Collins that $\w{H}=0 \Rightarrow
\omega \theta = 0$, we managed to prove the conjecture for the case
when $\textrm{div}\w{H} = 0$ (that is when the the magnetic part of the Weyl tensor is
solenoidal), with a $\gamma$-law equation of state\cite{norbetal1}. This result was further
generalised in a sense with our being able to establish the conjecture for
the case when $\textrm{div}\w{H} = 0$ and there is a sufficiently \emph{generic} equation of state\cite{carminatietal}. 
In the present article we will focus our attention on generalising
the result\cite{langth, carme} that the conjecture holds when the electric part of the Weyl tensor vanishes, by considering spacetimes for which the
electric part is solenoidal. Specifically, we prove

\subsubsection*{Theorem} Consider any shear-free perfect fluid solution of the Einstein field equations where the fluid pressure satisfies a barotropic 
equation of state and the spatial divergence of the electric part of the Weyl
tensor is zero. Then either the fluid is non-rotating or non-expanding.

\section{Notations and conventions}

\noindent We shall be examining shear-free, perfect fluid solutions of the Einstein field equations
\begin{equation}
R_{ab}-\frac12 Rg_{ab}=\mu u_{a}u_{b}+ph_{ab}, \label{eq1}
\end{equation}
\noindent where $\mathbf{u}$ is the future-pointing (time-like) unit tangent vector to the flow, $\mu$ and $p$ are the energy density and pressure of the fluid, respectively, and $h_{ab}$ $=g_{ab}+u_{a}u_{b}$ is the projection tensor into the rest space of the observers with 4-velocity $\mathbf{u}$. The vanishing of the shear can be
expressed by
\begin{equation}
u_{a;b}=\frac{1}{3}\theta h_{ab}+\omega_{ab}-\dot{u}_{a}u_{b}\ .
\label{eq2}
\end{equation}
We shall assume familiarity with the notation and conventions of the orthonormal tetrad formalism as given by MacCallum \cite{mal}. We begin the
analysis by choosing, what we believe to be, a well suited tetrad alignment for the physical problem at hand. This is the same as in our previous article \cite{norbetal1}. First, $\mathbf{e}_{0}$ and $\mathbf{e}_{3}$ are aligned with
$\mathbf{u}$ and $\w{\omega}$, respectively, such that $\w{\omega}$ $=\omega\mathbf{e}_{3}\neq 0$. The relevant variables then become $\mu,$ $p,$ $\theta,$ $\omega,$ $\dot{u}_{\alpha}$,
 $\Omega_{\alpha}$ together with the quantities $n_{\alpha\beta}$ and $a_{\alpha}$. Latin indices
will be tetrad indices taking the values 1, 2, 3, 4. Greek indices take the values 1, 2, 3, while uppercase Latin indices the values 1, 2 and any expressions involving these have to be 
read modulo 3 or 2 respectively. The sum of matter density and pressure will be written as
\begin{equation}
 \E=\mu+p .
\end{equation}
Secondly, it is always possible, using the remaining rotational freedom and making use of the Jacobi identities and field equations, to further specialize the tetrad so as to achieve%
\begin{eqnarray}
 \Omega_{1}=\Omega_{2}=\Omega_{3}+\omega=0,\label{eq3}\\
 n_{11}=n_{22}\equiv n.\nonumber
\end{eqnarray}
Herewith the tetrad is fixed up to rotations $\w{e}_1+i\w{e}_2 \to e^{i\alpha} (\w{e}_1+i\w{e}_2)$ satisfying $\partial_0\alpha =0$.
\noindent Because of computational advantages, we
will be replacing $n_{\alpha\beta}$ $(\alpha\neq\beta)$ and $a_{\alpha} $ with
the new variables $q_{\alpha}$ and $r_{\alpha}$ defined by $n_{\alpha
-1\,\alpha+1}$ $=(r_{\alpha}+q_{\alpha})/2$, and $a_{\alpha}$ $=(r_{\alpha
}-q_{\alpha})/2$. We will also
introduce extension variables, $z_{\alpha}$ and $j$ which are related to the
components of the spatial gradient of the expansion, by
\begin{eqnarray}\label{eq4}
\partial_{0}\theta=-\theta^{2}/3+2\omega^{2}-(\mu+3p)/2+j,\nonumber\\
\partial_{\alpha}\theta=z_{\alpha}, 
\end{eqnarray}
$j$ being the (3+1) covariant divergence of the acceleration,
\begin{equation}\label{defj}
j\equiv {\dot{u}^a}_{;a} =
\partial_{\alpha}\dot{u}^{\alpha}+\dot{u}^{\alpha}\dot{u}_{\alpha}-\dot{u}^{\alpha}
(r_{\alpha}-q_{\alpha}).
\end{equation}
In order to relate the components of $E_{ab}$ with the spatial
gradient of the acceleration, we will make use of the Ricci
identity~\cite{mart}
\begin{equation}\label{defE}
E_{\langle a b \rangle} = D_{\langle a }\dot{u}_{b\rangle}- \omega_{\langle
a}\omega_{b\rangle} + \dot{u}_{\langle a}\dot{u}_{b\rangle},
\end{equation}
where $S_{\langle a b \rangle}$ stands for the spatially projected
and trace-free part of a tensor $S_{ab}$.
\par The complete set of initial equations of the formalism are now the Einstein field equations and the Jacobi equations, which we present, using the simplifications above, in the appendix.
 First notice that the equations
(\ref{j8},\ref{j14}) immediately lead to evolution equations for
the variables $r_{\alpha}$ and $q_{\alpha}$,
\begin{eqnarray}
3 \partial_0 r_{\alpha}   = - z_{\alpha} - \theta (\dot{u}_{\alpha} + r_{\alpha}), \label{j8_14a}\\
3 \partial_0 q_{\alpha} = z_{\alpha} + \theta (\dot{u}_{\alpha}+q_{\alpha}), \label{j8_14b}
\end{eqnarray}
while (\ref{j1}) and the $(0\alpha)$ field equations
(\ref{ein01}-\ref{ein03}) give us the spatial derivatives of
$\omega$,
\begin{eqnarray}
\partial_1 \omega = \frac{2}{3} z_2-\omega(q_1+2 \dot{u}_1),\label{gradom1}\\
\partial_2 \omega = - \frac{2}{3} z_1+\omega (r_2-2\dot{u}_2),\label{gradom2}\\
\partial_3 \omega = \omega( \dot{u}_3+r_3-q_3),\label{gradom3}
\end{eqnarray}
together with the algebraic restriction
\begin{equation}\label{defn33}
n_{33} = \frac{2}{3\omega} z_3.
\end{equation}
The evolution equation for $n$ follows from (\ref{j11}):
\begin{equation}
\partial_0 n = -\frac{\theta}{3} n.
\end{equation}
Acting with the commutators $[\partial_0,\,
\partial_{\alpha}]$ and $[\partial_1,\, \partial_2]$  on the pressure and using (\ref{j7}) together with the conservation laws
\begin{equation}
\partial_0 \mu =-\E \theta ,\ \partial_{\alpha} p = -\E \dot{u}_{\alpha},\label{cons}
\end{equation}
leads to  a first set of evolution equations for the acceleration
and vorticity:
\begin{equation}\label{e0u3} \eqalign{
\partial_0 \dot{u}_{\alpha} = p' z_{\alpha} - G\theta \dot{u}_{\alpha},\\
\partial_0 \omega =\frac{1}{3} \omega \theta(-2+3p') ,}
\end{equation}
where we have defined 
\begin{equation}
G\equiv \frac{p''}{p'}\E -p'+\frac{1}{3}.
\end{equation}
The spatial derivatives of the acceleration can be obtained from
(\ref{defE}), using (\ref{defj}):
\begin{eqnarray}
\partial_{A}\dot{u}_{A} = -\frac{1}{3} \omega^2+\frac{1}{3} j-\dot{u}_{A+1} q_{A+1}+\dot{u}_{A-1} r_{A-1}-\dot{u}_A^2+E_{AA}\label{e1u1}\\
\partial_{3}\dot{u}_{3} = \frac{2}{3} \omega^2+\frac{1}{3} j-\dot{u}_1 q_1+\dot{u}_2 r_2-\dot{u}_3^2+E_{33}\label{u33}\\
\partial_1 \dot{u}_2 = -p'\omega \theta+q_2 \dot{u}_1+\frac{1}{2} n_{33} \dot{u}_3-\dot{u}_1 \dot{u}_2+E_{12}\label{e1u2}\\
\partial_2 \dot{u}_1 = p'\omega \theta-r_1 \dot{u}_2-\frac{1}{2} n_{33} \dot{u}_3-\dot{u}_1 \dot{u}_2+E_{12}\label{e2u1}\\
\partial_1 \dot{u}_3 = -\frac{1}{2} \dot{u}_2 n_{33}-r_3 \dot{u}_1-\dot{u}_1 \dot{u}_3+E_{13} \label{e1u3} \\
\partial_2 \dot{u}_3 = \frac{1}{2} \dot{u}_1 n_{33}+q_3 \dot{u}_2-\dot{u}_2 \dot{u}_3+E_{23} \label{e2u3} \\
\partial_3 \dot{u}_1 = -\frac{1}{2} \dot{u}_2 n_{33}+n \dot{u}_2+q_1 \dot{u}_3-\dot{u}_1 \dot{u}_3+E_{13}\\
\partial_3 \dot{u}_2 = \frac{1}{2} \dot{u}_1 n_{33}-n \dot{u}_1-r_2 \dot{u}_3-\dot{u}_2 \dot{u}_3+E_{23}.
\end{eqnarray}
Next we act with the $[\partial_0,\,
\partial_{\alpha}]$ commutators on $\omega$ and $\theta$ and use the propagation of (\ref{defn33}) along $\bi u$ in order to
obtain expressions for the evolution of $z_{\alpha}$ along $e_0$ and for the spatial gradient of $j$:
\begin{eqnarray}
\partial_0 z_1 &=& \theta (-1+p') z_1-\frac{1}{2}\omega(-1+9p') z_2+ \frac{1}{2}\theta\omega(9G-2)\dot{u}_2 \label{e0Z1}\\
\partial_0 z_2 &=&  \theta (-1+p') z_2+\frac{1}{2}\omega(-1+9p') z_1- \frac{1}{2}\theta\omega(9G-2)\dot{u}_1 \label{e0Z2} \\
\partial_0 z_3 &=& \theta (-1+p') z_3, \label{e0Z3}
\end{eqnarray}

\begin{eqnarray}
\partial_1 j &=& p'\theta z_1-\frac{1}{6} \omega (27p'+13)z_2+\frac{1}{3} \dot{u}_1 (18 \omega^2+\theta^2-3j-3\mu)-\frac{\E}{2p'}\dot{u}_1 
\nonumber \\
&+&\frac{1}{2}\theta \omega(9G-2) \dot{u}_2 +4 \omega^2 q_1 \label{e1j}\\
\partial_2 j &=& p'\theta z_2+\frac{1}{6} \omega (27p'+13)z_1+\frac{1}{3} \dot{u}_2 (18 \omega^2+\theta^2-3j-3\mu)-\frac{\E}{2p'}\dot{u}_2\nonumber \\
&-&\frac{1}{2}\theta \omega(9G-2) \dot{u}_1 -4 \omega^2 r_2 \label{e2j}\\
\partial_3 j &=& p'\theta z_3+\frac{1}{3} \dot{u}_3(\theta^2-18 \omega^2-3j-3\mu)-\frac{\E}{2p'} \dot{u}_3-4 (r_3-q_3) \omega^2 .
\end{eqnarray}
Now we may evaluate $\sum_{\alpha}[\partial_0,\,
\partial_{\alpha}]\dot{u}_{\alpha}$, using (\ref{defj}) and
(\ref{e1u1}-\ref{u33}), which leads to an expression for the evolution of $j$ in terms of $\partial_{\alpha} z_{\alpha}$.
\noindent With the aid of the field equations (\ref{ein11}-\ref{ein33}), their propagation along $\bi u$ and the use of the 
$[\partial_0,\, \partial_{\alpha}]$ commutators on $r_{\alpha}, q_{\alpha}, \dot{u}_{\alpha}$ (excluding $r_1,~q_2,~\dot{u}_3$ ) 
together with the $[\partial_1,\,\partial_2]$ commutators on $\omega$ and Jacobi equation (\ref{j4}) , one obtains algebraic 
expressions for the directional derivative $\partial_3 z_3$, as well as evolution equations for $j$ as
\begin{eqnarray}\label{e0jfirst}
\partial_0 j &=& \frac {\theta ( G'\E+p'-2Gp')}{p'}\dot{u}^2+(1-2G)(z_1\dot{u}_1+z_2\dot{u}_2+z_3\dot{u}_3)
\nonumber\\
&-&\frac{1}{3}\theta(1+3G)j- p'\theta (1-9p')\omega^2
\end{eqnarray}
\begin{eqnarray}\label{e3z_3}
\partial_3z_3&=&\frac{\theta(G'\E-2Gp')}{p'(1+3p')}(\dot{u}^2-3\dot{u}_3^2)-\frac{2G}{1+3p'}(z_1\dot{u}_1+z_2\dot{u}_2-2z_3\dot{u}_3)
-2z_3\dot{u}_3\nonumber \\
&+& \frac{2G+9p'^2-6p'+1}{1+3p'}\omega^2\theta+\frac{\theta(-2+3G)}{1+3p'}E_{33}+r_2z_2-q_1z_1
\end{eqnarray}
($G' \equiv \frac{dG}{d\mu}$ and $\dot{u}^2 \equiv \dot{u}_1^2+\dot{u}_2^2+\dot{u}_3^2 $).
Also the solutions of equations (\ref{j4}) and (\ref{ein12}) are given by
\begin{eqnarray}
\partial_2q_1&=&\frac{\theta(2Gp'-G'\E)}{3\omega p'(1+3p')}(\dot{u}^2-3\dot{u}_3^2)+\frac{2G}{3\omega(1+3p')}(z_1\dot{u}_1+z_2\dot{u}_2-2z_3\dot{u}_3)\nonumber\\
&+&\frac{r_3-3q_3+3\dot{u}_3}{3\omega}z_3-\frac{\theta(-2+3G)}{3\omega(1+3p')}E_{33}-\frac{9p'^2+2G-9p'}{3(1+3p')}\theta\omega\nonumber\\
&-&\frac{1}{3\omega}(r_2z_2-q_1z_1)+(r_1+q_1)r_2+n(q_3+r_3)-E_{12}
\end{eqnarray}
and 
\begin{eqnarray}
\partial_1r_2&=&\frac{\theta(2Gp'-G'\E)}{3\omega p'(1+3p')}(\dot{u}_1^2+\dot{u}_2^2-2\dot{u}_3^2)+\frac{2G}{3\omega(1+3p')}(z_1\dot{u}_1+z_2\dot{u}_2-2z_3\dot{u}_3)\nonumber\\
&+&\frac{3r_3-q_3+3\dot{u}_3}{3\omega}z_3-\frac{\theta(-2+3G)}{3\omega(1+3p')}E_{33}-\frac{9p'^2+2G-9p'}{3(1+3p')}\theta\omega\nonumber\\
&-&\frac{1}{3\omega}(r_2z_2-q_1z_1) -(r_2+q_2)q_1-n(q_3+r_3)+E_{12}.
\end{eqnarray}  
\noindent From the Ricci identity \cite{mart}, $H_{\langle a b \rangle}=2\dot{u}_{\langle a}\omega_{b\rangle}+D_{\langle a}\omega_{b\rangle}$, the components of the (trace-free) 
magnetic part of the Weyl curvature are given by
\begin{eqnarray}\label{def_H}
H_{11}&=&-\omega (\dot{u}_3+r_3), \, H_{22}=-\omega (\dot{u}_3-q_3) \nonumber \\
H_{13}&=& z_2/3 -\omega q_1, \, H_{23}= -z_1/3+\omega r_2, H_{12}=0.
\end{eqnarray}
At this stage we introduce the condition that $\w{E}$ is solenoidal, $\textrm{div}\w{E}=0$, which by the Bianchi identity\cite{mart}, 
$(\textrm{div}\w{E})_{a}=\frac{1}{3}D_a\mu-3\omega^bH_{ab}$ reduces to
\begin{eqnarray}
\E\dot{u}_1+3p'\omega(z_2-3\omega q_1) = 0,\label{divEcond1} \\
\E\dot{u}_2-3p'\omega(z_1-3\omega r_2) = 0,\label{divEcond2} \\
(\E+18\omega^2 p')\dot{u}_3-9\omega^2p'(-r_3+q_3) = 0. \label{divEcond3}
\end{eqnarray}
Apart from $\mu \omega \theta \E \neq 0$ we shall also assume  that $p'\neq 0, \pm 1/3, 1/9$, $\dot{u}_{1}^{2}+\dot{u}_{2}^{2}\neq0,$ 
$E_{ab}\neq0,$ $H_{ab}\neq0,\theta$ not just a function of either $\mu$ or $\omega,$ so as to avoid duplication of known results.
Propagating equations (\ref{divEcond1}) and (\ref{divEcond2}) gives then
\begin{eqnarray}
3p'[2\E-9(1-3p')\omega^2]z_A-\theta[2\E(6p'-2)+81G p'\omega^2]\dot{u}_A = 0,\label{EcondZ1}
\end{eqnarray}
These two equations show that $z_A$ is parallel to $\dot{u}_A$, unless  
\begin{equation}\label{eq16}
2\E-9(1-3p')\omega^2= 2\E(6p'-2)+81G p'\omega^2=0.
\end{equation}

\section{$\dot{u}_A$ parallel to $z_A$}
Taking the evolution of $\dot{u}_2z_1-\dot{u}_1z_2=0$ along $\w{u}$, this results in
\begin{eqnarray}
27p'(3G+3p'-1)\omega^2-\E(-1+27G p'+9p'-54p'^2)=0.\nonumber
\end{eqnarray}
If the coefficient of $\omega^2$ were zero, the latter equation would give \[\E(-1+9p')^2= 0.\]
Hence, the vorticity is a function of the matter density and one has $\partial_1\mu\partial_2\omega-\partial_2\omega\partial_1\mu=0$,
which reduces to $(\dot{u}_1z_1+\dot{u}_2z_2)\E=0$. It follows that $\dot{u}_1=\dot{u}_2=0$, which is impossible.

\section{$\dot{u}_A$ not parallel to $z_A$}
Eliminating $\omega$ from (\ref{eq16}) fixes the equation of state to be 
\begin{equation}\label{eqofstate}
9Gp'-18p'^2+9p'-1=0.
\end{equation} 
Note that a $\gamma$-law equation of state is then only possible when $p'=-1/3$ or $1/9$, so henceforth we will exclude 
all sub-cases in which $p'$ is constant.
By (\ref{eq16}) the vorticity is again a function of matter density. It follows that  
\begin{eqnarray}
\partial_A\mu\partial_3\omega-\partial_A\omega\partial_3\mu=0,\nonumber
\end{eqnarray}
which after eliminating $q_1,~r_2$ and $q_3$ using equations (\ref{divEcond1}), (\ref{divEcond2}) and (\ref{divEcond3}) yield
\begin{eqnarray}
\E(-3\omega\dot{u}_A+z_{A+1})\dot{u}_3=0.\nonumber
\end{eqnarray}
It follows that $\dot{u}_3$ is zero, as otherwise, propagation of equation (\ref{eq16}) along $e_A$ would imply $\E(-1+3p')\dot{u}_A=0$
and again $\dot{u}_A=0$.\\
When $\dot{u}_3$ is zero, the expressions (\ref{e0u3}), (\ref{e1u3}), (\ref{e2u3}), (\ref{u33}) and (\ref{defn33}) show that
\begin{eqnarray}
z_3=0, \ \ \ \ \ n_{33}=0, \ \ \ \ \ E_{13}=\dot{u}_1r_3, \ \ \ \ \ E_{23}=-\dot{u}_2q_3,\nonumber \\
E_{33}=-\frac{1}{3}(2\omega^2+j)+\dot{u}_1q_1-\dot{u}_2r_2. \nonumber
\end{eqnarray}
Moreover, the conditions $\textrm{div}\w{E}=0$, (\ref{divEcond1}-\ref{divEcond3}) can now be rewritten as follows
\begin{eqnarray}\label{Econd}\eqalign{
\E\dot{u}_1+3\omega p'(z_2-3\omega q_1)= 0,\\
\E\dot{u}_2-3\omega p'(z_1-3\omega r_2)= 0,\\
p'(q_3-r_3)\omega^2=0.}
\end{eqnarray}
Propagating equation (\ref{eq16})(a) along $\w{e}_1$ and $\w{e}_2$, using equations (\ref{Econd}), yields
\begin{eqnarray}
\E(-\dot{u}_1+3p'q_1)=0,\label{qr_12}\\
\E(\dot{u}_2+3p'r_2)=0\label{rq_21}.
\end{eqnarray}
Again propagating the latter equations 
along $\w{e}_1$ and $\w{e}_2$ respectively and eliminating $E_{12}$ shows then
\begin{eqnarray}\label{q1r2}
\fl 9\theta p'^2(3p'-1)(18p'^2-15p'+1)j+9p'^2(3p'-1)(36p'^2-9p'+5)(\dot{u}_1z_1+\dot{u}_2z_2)\nonumber\\
\fl \ \ \ \ \ \ -9\theta(3p'-1)(162p'^4+108p'^3-162p'^2+21p'-1)(\dot{u}_1^2+\dot{u}_2^2)\nonumber\\
\fl \ \ \ \ \ \ +6p'^3\theta(27p'^2-18p'+7)\E=0.
\end{eqnarray}
Now we use $z_3=0$ and equations (\ref{eq16}-\ref{qr_12}), (\ref{rq_21}) and (\ref{Econd}) to simplify equation (\ref{e3z_3} ) to
\begin{eqnarray}\label{z33}
\fl (3p'-1)(162p'^4+108p'^3-162p'^2+21p'-1)(\dot{u}_1^2+\dot{u}_2^2)-6(27p'^2-18p'+7)\E p'^3\nonumber\\
\fl \ \ \ \ \ \ -9p'^2(3p'-1)(18p'^2-15p'+1)j=0.
\end{eqnarray}
This can now  be combined with (\ref{q1r2}) to give
\begin{eqnarray}
p'^2(3p'-1)(36p'^2-9p'+5)(z_1\dot{u}_1+z_2\dot{u}_2)=0.\nonumber
\end{eqnarray}
It follows that $\dot{u}_A$ is orthogonal to $z_A$, after which propagation of equation (\ref{q1r2}) along $\w{e}_0$ gives 
\begin{eqnarray}\label{eq1_0}
 2(1296p'^4-2106p'^3+522p'^2+9p'-5)(3p'-1)^2(\dot{u}_1^2+\dot{u}_2^2)\nonumber\\
 \ \ \ \ \ \ -24p'^2\E(p'+1)(81p'^3-207p'^2+93p'-7)=0.
\end{eqnarray}
Taking the time derivative of the latter equation, eliminating $j, z_1$ using equation (\ref{q1r2}) and $z_2\dot{u}_2+z_1\dot{u}_1=0$, 
results in a new equation between $p,p'$ and $\dot{u}_1^2+\dot{u}_2^2$. Eliminating the acceleration between the latter pair eventually results in
\begin{eqnarray}
\frac{\E p'^3(3p'-1)(18p'^2-15p'+1)}{1296p'^4-2106p'^3+522p'^2+9p'-5}\times \nonumber \\
\fl (26244p'^6-59778p'^5-106515p'^4 +156249p'^3-43065p'^2+3267p'-26)&=&0,\nonumber
\end{eqnarray}
showing that $p'$ has to be a constant, in contradiction with $\ref{eqofstate}$ and our assumptions $p\neq-1/3, 1/9$.

\section{Discussion}
Newtonian homogeneous cosmologies are known\cite{senovilla} to provide counter-examples to the \emph{Newtonian} shear-free fluid conjecture.
In these models the spatial divergence of the tidal field (the Newtonian analogue of $\w{E}$) automatically vanishes, such that, 
when searching for possible counter-examples to the \emph{relativistic} shear-free fluid conjecture, it is tempting to look for 
candidates within the class of perfect fluids having a solenoidal electric part of the Weyl tensor. The result obtained in this paper shows
 that this attempt remarkably (or luckily ---depending on one's attitudes towards the conjecture) fails:  shear-free perfect fluids obeying a barotropic
 equation of state and having $\textrm{div}\w{E}=0$ are all non-rotating or non-expanding, in contrast to their Newtonian brethren.
Whether the conjecture in its full generality is true or false still remains an intriguing issue.
 
\section{Appendix A}

Commutator relations, using $\sigma_{\alpha \beta}=0$:
\begin{equation}
\eqalign{ \fl [\partial_0 , \partial_1] = {\dot u}_1 \partial_0
-\theta_1 \partial_1 +(\omega_3+\Omega_3) \partial_2
-(\omega_2+\Omega_2) \partial_3 \label{comms1}\\
\fl [\partial_0 , \partial_2] = {\dot u}_2 \partial_0
-(\omega_3+\Omega_3) \partial_1 -\theta_2 \partial_2
+(\omega_1+\Omega_1) \partial_3 \label{comms2}\\
\fl [\partial_0 , \partial_3] = {\dot u}_3 \partial_0
+(\omega_2+\Omega_2) \partial_1 -\theta_3 \partial_3
-(\omega_1+\Omega_1) \partial_2 \label{comms3}\\
\fl [\partial_1 , \partial_2] = -2 \omega_3 \partial_0 + q_2
\partial_1
+r_1 \partial_2 +n_{33} \partial_3 \label{comms4}\\
\fl [\partial_2 , \partial_3] =  -2 \omega_1 \partial_0 + q_3
\partial_2
+r_2 \partial_3 +n_{11} \partial_1 \label{comms5}\\
\fl [\partial_3 , \partial_1] = -2 \omega_2 \partial_0 + r_3
\partial_1 +q_1 \partial_3+n_{22} \partial_2 \label{comms6}}
\end{equation}\ \\

\noindent Using in addition the simplifications
$\omega_1=\omega_2=\Omega_1=\Omega_2=0$, $\Omega_3 = - \omega_3=-\omega$,
$n_{11}=n_{22}=n$ one obtains the:\\

\noindent Jacobi equations:
\begin{eqnarray}
\fl \partial_3 \omega  = \omega (\dot{u}_3+r_3-q_3) \label{j1} \\
\fl \partial_1 n + \partial_2 r_3  + \partial_3 q_2  - n (r_1 - q_1 ) - r_3 r_2 + q_3 q_2 = 0 \label{j2}\\
\fl \partial_2 n  + \partial_3 r_1  + \partial_1 q_3  - n (r_2 - n q_2) - r_3 r_1 + q_3 q_1  = 0 \label{j3}\\
\fl \partial_1 r_2 +\partial_2 q_1 +\partial_3 n_{33} -\frac{2}{3} \theta \omega-r_2 r_1+q_2 q_1-n_{33}( r_3- q_3) = 0 \label{j4}\\
\fl \partial_2 \dot{u}_3  - \partial_3 \dot{u}_2  - n \dot{u}_1 - q_3 \dot{u}_2  -  r_2 \dot{u}_3 = 0 \label{j5}\\
\fl \partial_3 \dot{u}_1  - \partial_1 \dot{u}_3  - n \dot{u}_2 - r_3 \dot{u}_1   - q_1 \dot{u}_3  = 0 \label{j6}\\
\fl 2 \partial_0 \omega +\partial_1 \dot{u}_2 -\partial_2 \dot{u}_1 -\dot{u}_1 q_2-\dot{u}_2 r_1- \dot{u}_3 n_{33}+\frac{4}{3} \theta \omega = 0 \label{j7}\\
\fl \partial_0 (r_{\alpha}-q_{\alpha})-\frac{1}{3}\partial_{\alpha} \theta +z_{\alpha}+\frac{\theta}{3}(r_{\alpha}-q_{\alpha}+2 \dot{u}_{\alpha}) = 0 \label{j8}\\
\fl 3 \partial_0 n  + 3 \partial_3 \omega  + n \theta - 3 \omega (\dot{u}_3 + r_3- q_3) = 0 \label{j11}\\
\fl 3 \partial_0 n_{33} + 3 \partial_3 \omega + n_{33} \theta - 3 \omega (\dot{u}_3  + r_3 - q_3) = 0 \label{j13}\\
\fl \partial_0
(r_{\alpha}+q_{\alpha})+\frac{1}{3}\theta(r_{\alpha}+q_{\alpha})
= 0 \label{j14}
\end{eqnarray}\ \\

\noindent Einstein equations:
\begin{eqnarray}
\fl \partial_0 \theta = -\frac{1}{3}\theta^2+2\omega^2-\frac{1}{2}(\mu+3 p)+j \label{ein00}\\
\fl 2/3 z_1 + \partial_2 \omega - \omega (r_2 - 2  \dot{u}_2) = 0\label{ein01}\\
\fl 2/3 z_2- \partial_1 \omega - \omega ( q_1 + 2  \dot{u}_1 ) = 0\label{ein02}\\
\fl 2/3 z_3- \omega n_{33} = 0\label{ein03}\\
\fl -\partial_1 r_2 +\partial_2 q_1 -r_1 r_2-q_1 q_2-2 r_2 q_1-2 r_3 n+n_{33} (r_3+q_3)-2 q_3 n =\nonumber \\
 -\partial_1 \dot{u}_2-\partial_2 \dot{u}_1-2 \dot{u}_1 \dot{u}_2+\dot{u}_1 q_2-\dot{u}_2 r_1\label{ein12}\\
\fl  -\partial_2 r_3 +\partial_3 q_2 +\partial_1 n -\partial_1 n_{33} -n (r_1-q_1)-2 q_1 n_{33}-r_2 r_3-q_2 q_3 -2 r_3 q_2= \nonumber \\
-\partial_2 \dot{u}_3-\partial_3 \dot{u}_2 +\dot{u}_1 (n_{33}-n) -2 \dot{u}_2 \dot{u}_3+\dot{u}_2 q_3-\dot{u}_3 r_2 \label{ein23}\\
\fl \partial_1 q_3 -\partial_3 r_1 +\partial_2 n_{33} -\partial_2 n -r_2 (2  n_{33}-n)-n q_2-r_1 r_3-q_1 q_3-2 r_1 q_3 =\nonumber  \\
-\partial_1 \dot{u}_3 -\partial_3 \dot{u}_1 -\dot{u}_1(2  \dot{u}_3+r_3)+\dot{u}_2(n- n_{33})+\dot{u}_3 q_1\label{ein13}\\
\fl -\partial_1 r_1 +\partial_1 q_1 +\partial_2 q_2 -\partial_3 r_3+q_2^2+r_3^2+\frac{1}{2} n_{33}^2-n n_{33}+r_1^2+q_1^2-r_2 q_2-r_3 q_3 = \nonumber \\
\frac{2}{9} \theta^2+\frac{8}{3} \omega^2-\frac{1}{3}(2 \mu- j)-\partial_1 \dot{u}_1-\dot{u}_1^2+\dot{u}_3 r_3-\dot{u}_2 q_2\label{ein11}\\
\fl -\partial_2 r_2+\partial_2 q_2-\partial_1 r_1+\partial_3 q_3 +q_3^2+r_1^2+\frac{1}{2} n_{33}^2-n n_{33}+r_2^2+q_2^2-r_1 q_1-r_3 q_3 =\nonumber  \\
\frac{2}{9} \theta^2+\frac{8}{3} \omega^2-\frac{1}{3}(2\mu-j)-\partial_2 \dot{u}_2-\dot{u}_2^2+\dot{u}_1 r_1-\dot{u}_3 q_3\label{ein22}\\
\fl -\partial_3 r_3 +\partial_3 q_3 +\partial_1 q_1 -\partial_2 r_2 +q_1^2+r_2^2-\frac{1}{2} n_{33}^2+r_3^2+q_3^2-r_1 q_1-r_2 q_2 = \nonumber \\
\frac{2}{9} \theta^2+\frac{1}{3}(2 \omega^2-2\mu+j)-\partial_3
\dot{u}_3 -\dot{u}_3^2+\dot{u}_2 r_2-\dot{u}_1 q_1\label{ein33}
\end{eqnarray}


\section*{References}

\begin{thebibliography}{99}

\bibitem {spat1} Banerji S (1968) Prog. Theor. Phys. \textbf{39} 365

\bibitem {carmn}Carminati J (1990) J. Math. Phys. \textbf{31 }2434


\bibitem {carm3a}Carminati J and Cyganowski S (1996) Class. Quantum Grav
\textbf{13} 1805

\bibitem {carm3b}Carminati J and Cyganowski S (1997) Class. Quantum Grav
\textbf{14} 1167

\bibitem {carminatietal} Carminati J, Karimian H R, Van den Bergh N and Vu K T  (2009) Class. Quantum Grav. \textbf{26} 195002

\bibitem {coley}Coley A A (1991) Class. Quantum Grav. \textbf{8 }955

\bibitem {collinsm} Collins C B (1984) J. Math. Phys. \textbf{25} 995

\bibitem {collinsKV} Collins C B (1986) Can. J. Phys. \textbf{64}, 191

\bibitem {carme} Cyganowski S and Carminati J (2000) Gen. Rel. Grav.
\textbf{32,} 221

\bibitem {np2}Czapor S R, McLenaghan R G and Carminati J (1992). Gen.
Rel. Grav. \textbf{24} 911

\bibitem {dust1}Ellis G F R (1967) J. Math. Phys. \textbf{8 }1171


\bibitem {dust2}G\"odel K (1949) Rev. Mod. Phys. \textbf{21} 447

\bibitem {spat2} King A R and Ellis G F R (1973) Commun. Math. Phys.
\textbf{31} 209

\bibitem {langth} Lang J M (1993) \emph{Contributions to the study of general relativistic
 shear-free perfect fluids} Ph. D. thesis, University of Waterloo, Canada

\bibitem {langcollins} Lang J M and Collins C B (1988) Gen. Rel.
Grav.\textbf{\ 20} 683

\bibitem {mart}Maartens R and Bassett B A (1998) Class. Quantum Grav. 15 705

\bibitem {mal}MacCallum M A H 1971 Cosmological Models from a Geometric Point
of View (Cargese) vol 6 (New York:Gordon and Breach) p 61


\bibitem {dust3}Sch\"ucking E (1957) Naturwissenschaften \textbf{19} 507

\bibitem {senovilla} Senovilla J M M, Sopuerta C F, Szekeres P (1998) Gen. rel. Grav. \textbf{30} 3

\bibitem {sopuerta}Sopuerta C F (1998) Class. Quantum Grav. \textbf{15} 1043

\bibitem {rad}Treciokas R and Ellis G F R (1971) Commun. Math.
Phys.\textbf{\ 23} 1

\bibitem {oframe}Van den Bergh N, (1988) Class. Quantum Grav. \textbf{5,} L169

\bibitem {norbetal1}Van den Bergh N, Carminati J and Karimian H R (2007)
Class. Quantum Grav. \textbf{24} 1

\bibitem {ghpII}Vu K T and Carminati J (2003). Gen. Rel.
Grav.\textbf{\ 35} 263

\bibitem {stem}Vu K T and Carminati J (2008). Preprint.

\bibitem {39state2}Van den Bergh N (1999) Class. Quantum Grav. \textbf{16} 117

\bibitem {collinp}White A J and Collins C B (1984) J. Math. Phys.
\textbf{25 }332

\end{thebibliography}
\end{document}